\documentclass[twoside,11pt]{article}

\usepackage[arxiv-submission]{melba}

\usepackage{amsmath,amsfonts}

\melbaheading{YYYY:NNN}{https://www.melba-journal.org/article/XX-AA}{2022}{1-?}{m1/yyyy}{m2/yyyy}{Singla et al}{}{}

\ShortHeadings{The Open Kidney Ultrasound Data Set}{Singla \textit{et al.}}
\firstpageno{1}

\title{The Open Kidney Ultrasound Data Set}

\author{\name Rohit Singla \email rsingla@ece.ubc.ca \\ 
        \addr Biomedical Engineering, University of British Columbia, Vancouver, BC, Canada
        \AND
        \name Cailin Ringstrom \email ceringstrom@ece.ubc.ca \\ 
        \addr Electrical and Computer Engineering, University of British Columbia, Vancouver, BC, Canada        
        \AND
        \name Grace Hu  \\ 
        \addr Electrical and Computer Engineering, University of British Columbia, Vancouver, BC, Canada         
        \AND
        \name Victoria Lessoway \\ 
        \addr Electrical and Computer Engineering, University of British Columbia, Vancouver, BC, Canada           
        \AND
        \name Janice Reid \\ 
        \addr Electrical and Computer Engineering, University of British Columbia, Vancouver, BC, Canada           
        \AND
        \name Christopher Nguan \email chris.nguan@ubcurology.com \\ 
        \addr Urologic Sciences, University of British Columbia, Vancouver, BC, Canada        
        \AND
        \name Robert Rohling \email rohling@ece.ubc.ca \\ 
        \addr Electrical and Computer Engineering and Mechanical Engineering, University of British Columbia, Vancouver, BC, Canada
}

\begin{document}

\maketitle

\begin{abstract}
	Ultrasound, because of its low cost, non-ionizing, and non-invasive characteristics, has established itself as a cornerstone radiological examination. Research on ultrasound applications has also expanded, especially with image analysis with machine learning. However, ultrasound data are frequently restricted to closed data sets, with only a few openly available. Despite being a frequently examined organ, the kidney lacks a publicly available ultrasonography data set. The proposed Open Kidney Ultrasound Data Set is the first publicly available set of kidney brightness mode (B-mode) ultrasound data that includes annotations for multi-class semantic segmentation. It is based on data retrospectively collected in a 5-year period from over 500 patients with a mean age of 53.2 ± 14.7 years, body mass index of 27.0 ± 5.4 kg/m2, and most common primary diseases being diabetes mellitus, immunoglobulin A (IgA) nephropathy, and hypertension. There are labels for the view and fine-grained manual annotations from two expert sonographers. Notably, this data includes native and transplanted kidneys. Initial bench-marking measurements are performed, demonstrating a state-of-the-art algorithm achieving a Dice Sorenson Coefficient of 0.85 for the kidney capsule. This data set is a high-quality data set, including two sets of expert annotations, with a larger breadth of images than previously available. In increasing access to kidney ultrasound data, future researchers may be able to create novel image analysis techniques for tissue characterization, disease detection, and prognostication.
\end{abstract}


\section{Introduction}
	Ultrasound remains a cornerstone of radiological assessment. \citep{szabo2004diagnostic} It is inexpensive, non-invasive, and real-time while using no ionizing radiation	\citep{szabo2004diagnostic}. Parallel advancements in technology have resulted in the availability of fully equipped devices in a portable smartphone-based form factor at a fraction of the cost of a larger cart-based system 	\citep{gerardo2018fabrication,rothberg2021ultrasound}. Ultrasound is still the primary line of examination for organs such as the heart, liver, and the kidney.\citep{remer2014acr,taffel2017acr} The American College of Radiologists recommends ultrasound as the first-line imaging modality for suspected kidney dysfunction in both native and transplanted kidneys.\citep{remer2014acr,taffel2017acr} This imaging modality thus plays a potential role in the clinical management of 1 in 10 adults globally who suffer from chronic kidney disease, representing over 800 million individuals.\citep{kovesdy2022epidemiology} When it comes to kidney pathology, changes in morphological characteristics (both increases and decreases) have been linked to disease severity.\citep{singla2022kidney} These include kidney length\citep{widjaja2004ultrasound,zanoli2014renal}, total kidney volume \citep{kim2008usefulness,sanusi2009relationship}, and cortical thickness \citep{korkmaz2018clinical}. These variables are frequently extracted manually by experts, resulting in intra- and inter-rater heterogeneity. For example, volume is approximated using ellipsoid-based equations with length-dimensional inputs rather than by careful segmentation of each ultrasound frame that contains the kidney. The latter, while more accurate, is a time-consuming and laborious procedure\citep{singla2022kidney}. Additionally, identifying the kidney's sections, notably the cortex and medulla, requires considerable expertise. There is therefore a need for objective automatic analysis for segmentation and measurement. Kidney ultrasound image analysis can be split into two segmentation tasks, either of which can be used for subsequent downstream tasks such as regional feature extraction. The first challenge is to extract the renal capsule's (border) binary segmentation from the background ultrasound image. This is useful for determining whether the complete kidney is visible and can be used to determine the size of the kidney. The second aim is to perform a semantic segmentation of specific areas of the kidney, such as the cortex and medulla, using a multi-class approach. These regions are difficult for the novice to interpret visually, and changes in their size and echogenicity of these regions also have clinical importance as well.\citep{singla2022kidney}

\section{Related Works}
	Machine learning has shown remarkable performance in tackling difficult semantic segmentation tasks and is a promising solution to address these two tasks. However, access to high-quality medical imaging data continues to be a challenge. Before data can be made broadly accessible, extensive authorization processes requiring institutional and research ethics board approvals, anonymization of retrieved images, and privacy considerations must be implemented. The medical imaging community's recent efforts have expanded the accessibility of this data. The Medical Segmentation Decathlon, a seminal moment in the development of publicly available medical data, focuses on ten distinct organ data sets with multiple medical imaging modalities \citep{antonelli2022medical}. However, at present, the Decathlon does not include ultrasound imaging. Alternative ultrasound data sets exist for the breast \citep{al2020dataset}, thyroid \citep{zhou2020thyroid}, heart \citep{ouyang2020video}, and lung \citep{born2021accelerating}. To our knowledge, no data set exists for kidney ultrasonography. Meanwhile, the contemporary kidney ultrasonography segmentation literature frequently reports on limited sample sizes, makes use of proprietary data sets that are inaccessible to other researchers, or entirely omits transplanted kidneys \citep{xie2005segmentation, dahdouh2009real, jokar2013kidney, mendoza2013kidney,marsousi2016automated, torres2018kidney,yin2020automatic}. Additionally, there is inconsistency in the evaluation of reported algorithms. This exacerbates the growing reproducibility dilemma in medical imaging machine learning, as demonstrated by image analysis challenges\citep{maier2020bias}. There exists a need for a reproducible data of kidney ultrasound images with detailed expert annotations and thorough data set characterization. 
	
	To meet this need, the Open Kidney Ultrasound Data Set is presented. This contribution includes 514 anonymized two-dimensional B-mode abdominal ultrasound images from the same number of patients, as well as two separate sets of manually generated fine-grained polygon annotations from two expert sonographers using a systematic protocol developed and agreed upon in advance. Labels for type of view and type of kidney are also included. Additionally, we present benchmark findings utilizing a state-of-the-art segmentation neural network. This will enable us to further enhance scholarly research in ultrasound machine learning. This data set may help standardise ultrasound segmentation benchmarking and, in the long term, may help reduce ultrasound interpretation efforts while significantly simplifying ultrasound use.

\section{Methods}
	\subsection{Image Acquisition}
	The retrospective collection of the B-mode ultrasound images was approved by our institution’s Research Ethics Board (H21-02375). The ultrasound images were originally acquired between January 2015 and September 2019. These B-mode ultrasound images are collected from real patients who had a clinical indication to receive an ultrasound investigation of their kidneys. Consequently, a significant portion are obtained at the bedside or intensive care units, rather than finely controlled laboratory conditions. The participant population includes adults with chronic kidney disease, prospective kidney donors, and adults with a transplanted kidney. A variety of ultrasound manufacturers and machine models are represented. This includes the ACUSON Sequoia with 6C2 and 4V1 transducers, the General Electric LOGIQ E10 with C1-6 and C2-9 transducers, the Philips Affiniti 70G with C5-1 and C9-2 transducers, the Philips Epiq 5G with C5-1 and C9-2 transducers, the Philips Epiq 7G with C5-1 and C9-2 transducers, the Philips iU22 with the C5-1 transducer, the Siemens S2000 with 4V1 and 6C1 transducers, and the Toshiba TUS-A500 with a PVT-375BT transducer.

    \subsection{Pre-processing}
    Full ultrasound videos (termed cines) were retrieved with the Digital Imaging and Communications in Medicine (DICOM) protocol directly from hospital picture archiving and communication systems. Anonymization included the removal of DICOM metadata, removal of any personally identifiable information from the file, and masking of any personally identifiable information on the images. Images were manually reviewed to confirm anonymization. From each cine, a random image was sampled from each one to ensure various cross-sections of the kidney were obtained. Files not containing the kidney were manually removed. The files were then converted to Portable Network Graphics (PNG) format for annotation using the Visual Geometry Group Image Annotator (VIA) software \citep{dutta2019via}. The anonymized PNG files were used for annotation. After annotation, additional graphics, and illustrations (ex: image settings or study date) on the ultrasound images were automatically removed. Images are provided in their original dimensionality.

    \subsection{Annotations}
    Two sonographers (V.L. and J.R.) manually annotated the kidney images. Both sonographers are registered generalist sonographers as well as registered cardiac sonographers. They remain in good standing with Sonography Canada and have $\geq$ 30 years of experience performing and teaching ultrasound imaging. Both have previous experience performing manual annotations for ultrasound, both in teaching and research settings.
    
    Through an iterative and collaborative process, the annotation goals, definitions, and guidelines were first established prior to annotations performed. Sonographers were asked to perform three tasks for each image: a) assess quality, b) label the view, c) annotate the classes. A random sampling of 20 kidney ultrasound images were provided to the sonographers as a practice set. Upon review, the categories of unacceptable, poor, fair, and good were determined for quality. The view label indicates in which orientation relative to the kidney the image was acquired, with standard views consisting of the transverse and longitudinal plane of imaging. A third category, "other", was used for images acquired in a non-standard manner such as in an oblique plane. 
    
    Four classes for segmentation were defined as the kidney capsule, cortex, medulla, and central echogenic complex (CEC). The functional anatomy of the kidney is compromised of the renal parenchyma and the renal pelvis \citep{netter2018atlas}. The parenchyma consists of the cortex (the outermost portion of the kidney) and the medulla\citep{netter2018atlas}. The renal pelvis however is more complex, consisting of minor and major calyces, blood vessels, and fat\citep{netter2018atlas}. As these components are not able to be individually delineated in these images by the sonographers, the term CEC is introduced to aggregate the different anatomy into one class for segmentation.
    
    The annotations for each segmentation class are hand-drawn closed polygons. No instructions were given for the number of vertices to use, or guidelines on minimum size requirements. The practice set of 20 images were annotated by each sonographer for quality, view, and the segmentation classes. Annotations were reviewed collectively to achieve agreement on quality and view, identify limitations in annotations, and provide clarifications on how to annotate. The 20 images used are not included in the released data set. The guidelines established are available online.
    
    In the released data, the sonographers both agreed upon the quality of each image. They individually then labelled views and performed annotations. Unbeknownst to the sonographers, a random set of 10 images were repeated three times (20 copies of original data) throughout the data to ascertain intra-rater variability. The annotations are not aggregated or fused in any manner. The released data includes each individual sonographer’s annotations.

    \subsection{Quality Assurance}
    Two biomedical engineers (R.S. and C.R.) reviewed all images with annotations. Any errors such as a missing capsule boundary or repeated class label were identified and returned for correction. In images where there was a discrepancy in the presence of the cortex and medulla (i.e., having one class but not the other) were flagged. Sonographers were asked to review these images and achieve consensus on the correction needed. At the time of release, a date-stamped list of errata is maintained and is available online should further errors be identified. The errata is available at \url{github.com/rsingla92/kidneyUS/blob/main/README.md\#errata}
    
    \subsection{Imaging Data}
\begin{figure}
	    \centering
	    \includegraphics[scale=0.67]{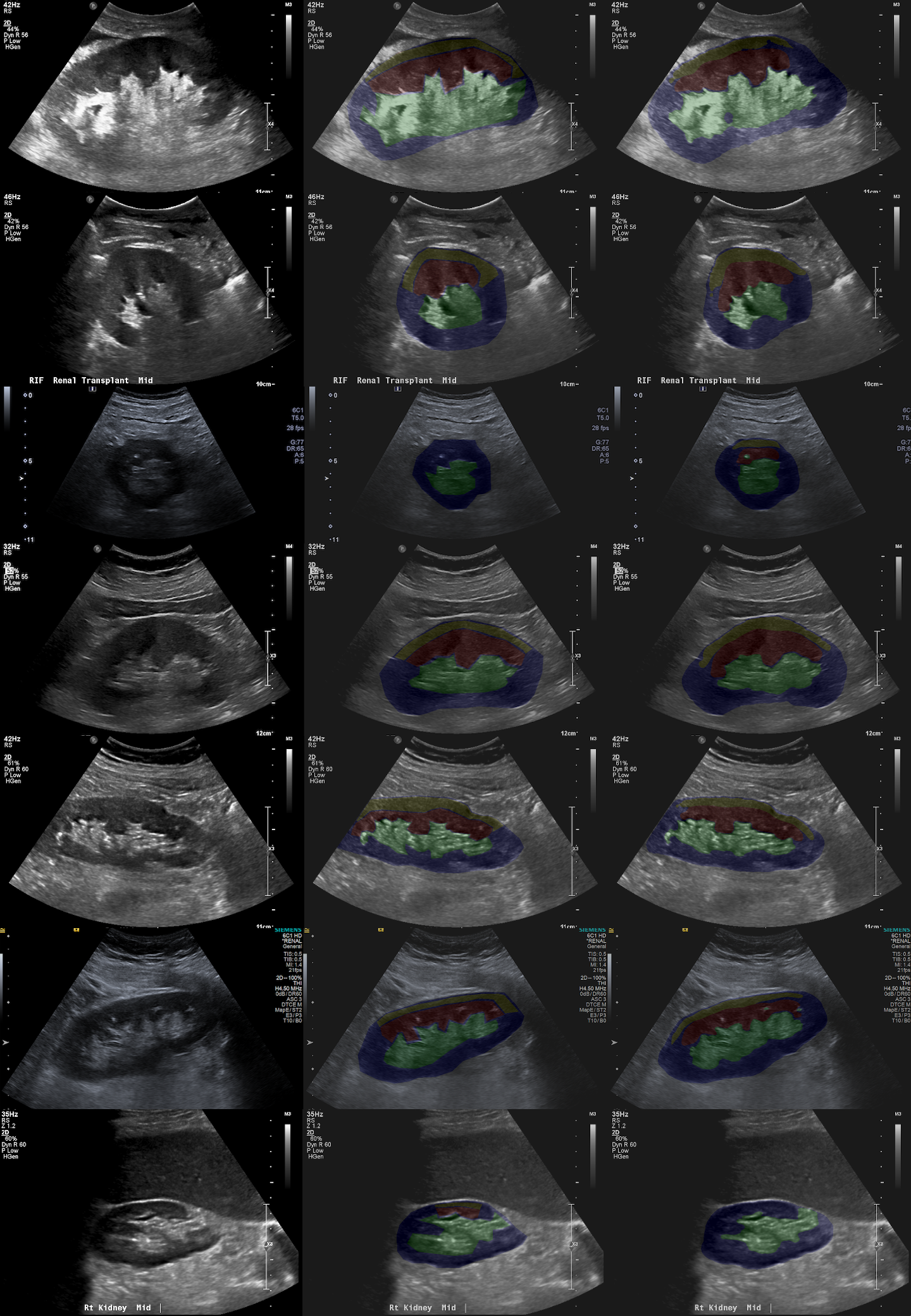}
	    \caption{Example images. (L to R columns): original image, the sonographer's annotation, and the predicted segmentations. Capsule in blue, cortex in yellow, medulla in red and green in CEC. Transplant kidneys are in odd rows, native ones in even rows.}
	    \label{fig:kidney_tiled}
	\end{figure}

    The Open Kidney Ultrasound Data Set is available at \url{https://rsingla92.github.io/kidneyUS/}. The data itself is shared via our institution’s instance of Microsoft OneDrive (Microsoft, Redmond, WA) and is an overall size of approximately 130 megabytes. It consists of 514 PNG files, each corresponding to a distinct patient. The additional 20 copied images used for validation and a list of these copied images are included. Each sonographer’s annotations are provided in a comma separated value (CSV) file, following the structure provided by VIA. Example images are seen in Fig. \ref{fig:kidney_tiled}.
    
    Due to privacy considerations and the risk of identification, patient-level demographic details are not available for the public. Instead, aggregated summary statistics of the patient population involved are provided. The mean (± standard deviation) of the patient population included is 53.2 ± 14.7 years, with 63\% of patients being males and 37\% being females. The mean (± standard deviation) body mass index was 27.0 ± 5.4 kg/m2. When examining the most recent estimated glomerular filtration rate (eGFR) to the time of imaging, the mean (± standard deviation) eGFR was 28.7 ± 21.3. The range of primary diagnoses causing end-stage kidney disease included 32\% due to type 1 or type 2 diabetes mellitus, 11\% due to immunoglobulin A (IgA) nephropathy, 10\% due to hypertension, and the remaining 47\% due to a myriad of other diseases which individually constituted $\leq 10$\% of the data. For the image data, Table 1 summarizes the image quality, frequency of views, kidney types, frequency of each class per frame, and the average size of annotation per class. Table 2 captures the number of vertices, the absolute pixel coverage of each class to the overall image, and the relative pixel coverage of the kidney’s compartments to the capsule in each set of annotations from the sonographers.

	\begin{table}[htb] 
		\centering
		\caption{Summary of quality, view, kidney type, and class frequency in terms of number of frames. The variance in class frequency is attributed to differences in sonographer interpretation.}
		\begin{tabular}{lr}
			\textbf{Quality (no. of frames)} & \\
			\hline
			Good & 1 (0.2\%)  \\
			Fair & 486 (91.0\%) \\ 
			Poor & 41 (7.7\%) \\
			Unsatisfactory & 6 (1.1\%) \\
			\hline
			\textbf{View (no. of frames)} & \\
			\hline
			Transverse & 145 (27.2\%)  \\
			Longitudinal & 371 (69.5\%) \\ 
			Other & 18 (3.4\%) \\
			\hline
			\textbf{Kidney Type (no. of frames)} & \\
			\hline
			Native & 148 (27.8\%)  \\
			Transplanted & 386 (72.2\%) \\ 
			\hline
			\textbf{Class (no. of frames)} & \\
			\hline
			Capsule & 452 ± 1\\
			Cortex & 452 ± 1 \\ 
			Medulla & 316 ± 4 \\
			Central Echogenic Complex & 315 ± 5 \\
			\hline
			\textbf{Average Pixels per Class (per 1000 pixels)} & \\
			\hline
			Capsule & 237 ± 7.7\\
			Cortex & 65.1 ± 1.3 \\ 
			Medulla & 35.2 ± 0.2 \\
			Central Echogenic Complex & 21.2 ± 0.6 \\
			\hline
		\end{tabular}
	\end{table}

	\begin{table}[htb] 
		\centering
		\caption{Summary of quality, view, kidney type, and class frequency in terms of number of frames. The variance in class frequency is attributed to differences in sonographer interpretation. Px: pixel, S1: sonographer 1, S2: sonographer 2.}
		\begin{tabular}{lll}
		     & \textbf{S1} & \textbf{S2} \\ 
			\hline
			\textbf{No. vertices (mean ± standard deviation)} \\
			Capsule & 18.9 ± 8.0 & 20.2 ± 8.0\\
			Cortex & 16.7 ± 15.2 & 15.3 ± 14.4 \\
			Medulla & 16.5 ± 15.5 & 21.5 ± 20.7 \\
			Central Echogenic Complex & 21.7 ± 13.5 & 31.3 ± 17.2 \\
			\hline
			\textbf{Absolute Px Coverage (median \%, [min, max]}) \\
			Capsule & 26.6 [0, 87.7] & 29.2 [0,93]\\
			Cortex & 1.8 [0, 19.6] & 1.7 [0, 19.6] \\
			Medulla & 3.1 [0, 28.8] & 2.9 [0, 26.0] \\
			Central Echogenic Complex & 6.8 [0, 30.6] & 7.3 [0, 32.2] \\
			\hline
			\textbf{Relative Px Coverage (median \%, [min, max}]) \\
			Cortex & 7.3 [0, 44.5] & 6.7 [0,65.3]\\
			Medulla & 13.9 [0, 48.4] & 12.6 [0, 77.6] \\
			Central Echogenic Complex  & 26.3 [0, 64.0] & 25.3 [0, 54.4] \\
			\hline
		\end{tabular}
	\end{table}
	
    \section{Technical Validation}
    Intra-rater variability was performed with each sonographer annotation 10 images three times in a blinded manner. Inter-rater variability was performed by randomly selecting 50 images and comparing the two in terms of Dice Sorenson Coefficient (DSC)\citep{reinke2021common}.
    
    \begin{table}[htb] 
		\centering
		\caption{Intra-rater and inter-rater variability between sonographers as measured using the Dice Sorenson Coefficient. CEC: Central echogenic complex.}
		\begin{tabular}{lllll}
		     & \textbf{Capsule} & \textbf{Cortex} & \textbf{Medulla} & \textbf{CEC}\\ 
			\hline
			\textbf{Intra-Rater} \\
			Sonographer 1 & 0.95 ± 0.02 & 0.62 ± 0.26 & 0.75 ± 0.21 & 0.85 ± 0.07\\
			Sonographer 2 & 0.96 ± 0.01 & 0.72 ± 0.20 & 0.80 ± 0.18 & 0.89 ± 0.06 \\
			\hline
			\textbf{Inter-Rater} & 0.93 ± 0.10 & 0.48 ± 0.37 & 0.54 ± 0.38 & 0.82 ± 0.17 \\
			\hline
		\end{tabular}
	\end{table}
		
    In the released data set, we include pre-trained models to provide initial benchmarks for the segmentation tasks so that users may reproduce our reported results. The recent release of nnU-net by \cite{isensee2021nnu} represents the current state of the art in medical image segmentation. Their fully automatic, self-configuring, out-of-the-box solution uses a data-adaptive variant of U-net that has been shown to generate state-of-the-art results in numerous image segmentation challenges \citep{isensee2021nnu}. Their approach uses a combination of fixed parameterization, heuristics, and empirical measures. Over 50 different tasks were included in their paper for segmentation. We use nnU-net to establish results for the binary segmentation of the kidney capsule and the multi-class segmentation of the kidney compartments.
    
    The data set of annotated kidney images was split into a training set and test set with an 80:20 ratio. Two-dimensional binary masks were created from the capsule annotation and multi-class masks were created from the annotation for the kidney compartments. The images and masks were converted from PNG to the Neuroimaging Informatics Technology Initiative (NifTI) file format. Then the predefined pre-processing and training procedures of the nnU-net were used which are based on heuristics and data set characteristics. The network was trained on five folds of the training data, and an ensemble of these was used for inference. Each network was trained for 500 epochs with a batch size of 4. The network is trained with a learning rate of 0.01 annealed throughout training, using the Adam optimizer, and a combined loss of DSC and cross-entropy loss. The training for each network took approximately 9 hours. The network was trained on the University of British Columbia Advanced Research Computing Sockeye platform using one 32GB Nvidia Tesla V100 GPU.
    
    Three versions of nnU-net were trained.  The first and second versions use annotations from only one of the two sonographers, creating sonographer-specific models. The third version randomly sampled annotations from both sonographers. For a given image, there was an equal (50\%) random chance of a sonographer’s annotation being selected. Pre-trained weights are made available.
    
	\begin{table}[htb] 
		\centering
		\caption{Segmentation validation using nnU-net. Network accuracies using annotations from models trained on individual sonographers and a random even sampling from both is reported. Hausdorff distance is reported in millimeters. CEC: central echogenic complex. DSC: Dice Sorenson Coefficient. HD: Hausdorff Distance. S1: Sonographer 1. S2: Sonographer 2.}
		\begin{tabular}{lllllll}
		     & & \textbf{Capsule} & \textbf{Cortex} & \textbf{Medulla} & \textbf{CEC} & \textbf{Mean}\\
			\hline
			\textbf{DSC} &	S1 Model & 0.87 & 0.43  & 0.48 & 0.76 & 0.63 \\
		                 &	S2 Model & 0.85 & 0.51  & 0.57 & 0.66 & 0.67 \\
                         &  Both     & 0.85 & 0.52  & 0.59 & 0.78 & 0.69 \\
			\hline
			\textbf{HD}  & S1 Model & 11.0 & 16.4 & 11.8 & 10.7 & 12.5 \\
			             & S2 Model & 15.5 & 13.7 & 9.3 & 15.9 & 13.6 \\
                         & Both          & 12.2 & 12.9 & 9.8 & 10.8 & 11.4 \\	
            \hline
		\end{tabular}
	\end{table}
	
	\begin{table}[htb] 
		\centering
		\caption{Sensitivity analysis of both the sonographer’s annotations and the model’s predictions given an erosion or dilation in each class’s annotation. Values are the absolute percent change in DSC.}
		\begin{tabular}{lllll}
		     Modification & & \textbf{Cortex} & \textbf{Medulla} & \textbf{CEC} \\
			\hline
			\textbf{1px erosion} &	Sonographer 1 & 3.9 & 2.0 & 2.5 \\
		                 &	Sonographer 2 & 5.2 & 2.8 & 2.4 \\
                         &	Model 1 & 7.2 & 5.8 & 0.8 \\
                         &	Model 2 & 7.0 & 3.3 & 1.4 \\
            \hline
           \textbf{10px erosion} &	Sonographer 1 & 46.8 & 26.1 & 28.4 \\
		                 &	Sonographer 2 & 62.3 & 36.2 & 27.2 \\
                         &	Model 1 & 37.7 & 29.9 & 16.4 \\
                         &	Model 2 & 44.1 & 27.0 & 15.9 \\
               \hline          
            \textbf{1px dilation} &	Sonographer 1 & 3.4 & 1.9 & 2.4 \\
		                 &	Sonographer 2 & 4.5 & 2.7 & 2.3 \\
                         &	Model 1 & 6.7 & 6.2 & 0.8 \\
                         &	Model 2 & 6.5 & 3.1 & 1.4 \\
              \hline           
            \textbf{10px dilation} &	Sonographer 1 & 22.5 & 14.8 & 19.2 \\
		                 &	Sonographer 2 & 29.7 & 20.5 & 18.8 \\
                         &	Model 1 & 42.5 & 55.1 & 9.0 \\
                         &	Model 2 & 44.9 & 18.9 & 14.3 \\
            \hline
		\end{tabular}
	\end{table}

Given the observed lower DSC values for the smaller regions of the cortex, medulla, and CEC, we performed a sensitivity analysis. For each of the sonographer’s original annotation, the polygons were eroded by 1 pixel, eroded by 10 pixels, dilated by 1 pixel, and dilated by 10 pixels. Similarly, for the two models based on individual sonographers, the same erosion/dilation was performed. The mean (± standard deviation) absolute percent change in DSC value for each class is reported. With pixel-to-millimetre scale factors being on the order of 0.1 mm for abdominal imaging, it can be observed that a 1 mm error (10 pixels) in prediction can cause significant change in the DSC. This demonstrates the challenges of using the DSC as an accuracy metric for these small regions, given its sensitivity to size and shape unawareness \citep{reinke2021common}

\subsection{Limitations and Sources of Error}
The generalizability of this data is limited as it is only obtained from one urban tertiary hospital. This is improved by the fact that this institution services individuals across the entire province of British Columbia, however inherent geographic disparities may still exist. Historical data from our institution also indicates that transplant recipients are predominantly Caucasian or Asian males.
    
    While the cortex and medulla are established terms, their anatomical delineation is unclear. Furthermore, there is no prior definition of the central echogenic complex. Our definition broadly incorporates different minor parts of the kidney’s anatomy. While the expert sonographers are experts in ultrasound, there many still exist discrepancies between their annotations and how another expert may interpret the image. This aleatoric uncertainty is captured to a small degree in the inter-rater variance above. Potential sources of error may include inaccurate segmentation masks or incorrect class labels. This has been mitigated through the quality assurance measures performed. 
    
    We also make no distinction between the importance of each class, and hence no specific weighting. The importance of a class would depend on the desired downstream task. The kidney capsule alone for example can be used to inform kidney bio-metrics such as size and shape of the organ. In other instances, the regional analysis of the cortex may be the most interesting as cortical thinning has been shown to be a prognostic factor of disease. Furthermore, being able to distinguish between cortex and medulla my be of clinical interest. This is because the inability to discriminate them in ultrasound, termed the loss of corticomedullary differentiation, is yet another indication of disease. Weighting different classes included in this work may result in class-specific or task-specific improvements.
    
    Furthermore, there are numerous potential exogenous factors which may impact performance of machine learning techniques. Consider, as one example, the role of 'acquisition shift' within medical imaging \citep{castro2020causality}. In using different devices or imaging acquisition conditions, scanner- or transducer-specific biases may be introduced. In \citeauthor{glocker2019machine}, the authors demonstrated site-specific scanners had subtle and significant effects on machine learning accuracy and generalization. With this consideration in mind, methods trained with the Open Kidney Data Set may face difficulties generalizing to ultrasound transducers not represented in the data; likewise methods trained on alternative data sources and tested on our data may observe a drop of performance.

    \subsection{Usage Notes}
    The data and code that are made available are under the Creative Commons Attribution Non-commercial Share Alike (CC BY-NC-SA) license. Data may not be used for commercial purposes. Due to accessibility and privacy terms, registration is required for manual verification prior to the release of data. 
    
    \subsection{Code Availability}
    Relevant code for masking, cropping data, reading, and processing summary statistics of labels, pre-trained models and additional code is available at: \url{https://github.com/rsingla92/kidneyUS}
    
    \section{Conclusion}
    In this work, we presented the Open Kidney Ultrasound Data Set, a high quality and well characterized data set of kidney ultrasound images with two sets of expert annotations. State-of-the-art models are trained and evaluated, yielding excellent segmentation accuracy on the largest class of the capsule with moderate accuracies on the more challenging inner classes. Sensitivity analysis shows high sensitivity to small errors in segmentations. In increasing access to kidney ultrasound data, future researchers may be able to create novel image analysis techniques for tissue characterization, disease detection, and prognostication.


\acks{The authors thank the patients whose data comprises this data set as well as the Biomedical Imaging and Artificial Intelligence Cluster for infrastructure and support. R.S. acknowledges funding from the Vanier Graduate Scholarship, the Kidney Foundation of Canada, and the American Society of Transplant Surgeons. C.R. acknowledges funding from Natural Sciences and Engineering Council of Canada (NSERC).}

%
\ethics{The work follows appropriate ethical standards in conducting research and writing the manuscript, following all applicable laws and regulations regarding treatment of animals or human subjects. Institutional research ethics board approval was obtained (H21-02375).}

\coi{RR and CN are shareholders and advisors to SonicIncytes, an ultrasound elastography imaging company. RR is co-investigator on grants supported by Philips Healthcare, Clarius Mobile Health Corp, and Change Healthcare. All other authors declare no conflicts of interest.}

\bibliography{sample}

\begin{thebibliography}{31}
\providecommand{\natexlab}[1]{#1}
\providecommand{\url}[1]{\texttt{#1}}
\expandafter\ifx\csname urlstyle\endcsname\relax
  \providecommand{\doi}[1]{doi: #1}\else
  \providecommand{\doi}{doi: \begingroup \urlstyle{rm}\Url}\fi

\bibitem[Al-Dhabyani et~al.(2020)Al-Dhabyani, Gomaa, Khaled, and
  Fahmy]{al2020dataset}
Walid Al-Dhabyani, Mohammed Gomaa, Hussien Khaled, and Aly Fahmy.
\newblock Dataset of breast ultrasound images.
\newblock \emph{Data in brief}, 28:\penalty0 104863, 2020.

\bibitem[Antonelli et~al.(2022)Antonelli, Reinke, Bakas, Farahani,
  Kopp-Schneider, Landman, Litjens, Menze, Ronneberger, Summers,
  et~al.]{antonelli2022medical}
Michela Antonelli, Annika Reinke, Spyridon Bakas, Keyvan Farahani, Annette
  Kopp-Schneider, Bennett~A Landman, Geert Litjens, Bjoern Menze, Olaf
  Ronneberger, Ronald~M Summers, et~al.
\newblock The medical segmentation decathlon.
\newblock \emph{Nature Communications}, 13\penalty0 (1):\penalty0 1--13, 2022.

\bibitem[Born et~al.(2021)Born, Wiedemann, Cossio, Buhre, Br{\"a}ndle,
  Leidermann, Goulet, Aujayeb, Moor, Rieck, et~al.]{born2021accelerating}
Jannis Born, Nina Wiedemann, Manuel Cossio, Charlotte Buhre, Gabriel
  Br{\"a}ndle, Konstantin Leidermann, Julie Goulet, Avinash Aujayeb, Michael
  Moor, Bastian Rieck, et~al.
\newblock Accelerating detection of lung pathologies with explainable
  ultrasound image analysis.
\newblock \emph{Applied Sciences}, 11\penalty0 (2):\penalty0 672, 2021.

\bibitem[Castro et~al.(2020)Castro, Walker, and Glocker]{castro2020causality}
Daniel~C Castro, Ian Walker, and Ben Glocker.
\newblock Causality matters in medical imaging.
\newblock \emph{Nature Communications}, 11\penalty0 (1):\penalty0 1--10, 2020.

\bibitem[Dahdouh et~al.(2009)Dahdouh, Frenoux, and Osorio]{dahdouh2009real}
S~Dahdouh, E~Frenoux, and A~Osorio.
\newblock Real-time kidney ultrasound image segmentation: a prospective study.
\newblock In \emph{Medical Imaging 2009: Ultrasonic Imaging and Signal
  Processing}, volume 7265, pages 134--142. SPIE, 2009.

\bibitem[Dutta and Zisserman(2019)]{dutta2019via}
Abhishek Dutta and Andrew Zisserman.
\newblock The via annotation software for images, audio and video.
\newblock In \emph{Proceedings of the 27th ACM international conference on
  multimedia}, pages 2276--2279, 2019.

\bibitem[Gerardo et~al.(2018)Gerardo, Cretu, and
  Rohling]{gerardo2018fabrication}
Carlos~D Gerardo, Edmond Cretu, and Robert Rohling.
\newblock Fabrication and testing of polymer-based capacitive micromachined
  ultrasound transducers for medical imaging.
\newblock \emph{Microsystems \& Nanoengineering}, 4\penalty0 (1):\penalty0
  1--12, 2018.

\bibitem[Glocker et~al.(2019)Glocker, Robinson, Castro, Dou, and
  Konukoglu]{glocker2019machine}
Ben Glocker, Robert Robinson, Daniel~C Castro, Qi~Dou, and Ender Konukoglu.
\newblock Machine learning with multi-site imaging data: an empirical study on
  the impact of scanner effects.
\newblock \emph{arXiv preprint arXiv:1910.04597}, 2019.

\bibitem[Isensee et~al.(2021)Isensee, Jaeger, Kohl, Petersen, and
  Maier-Hein]{isensee2021nnu}
Fabian Isensee, Paul~F Jaeger, Simon~AA Kohl, Jens Petersen, and Klaus~H
  Maier-Hein.
\newblock nnu-net: a self-configuring method for deep learning-based biomedical
  image segmentation.
\newblock \emph{Nature methods}, 18\penalty0 (2):\penalty0 203--211, 2021.

\bibitem[Jokar and Pourghassem(2013)]{jokar2013kidney}
Ehsan Jokar and Hossein Pourghassem.
\newblock Kidney segmentation in ultrasound images using curvelet transform and
  shape prior.
\newblock In \emph{2013 International Conference on Communication Systems and
  Network Technologies}, pages 180--185. IEEE, 2013.

\bibitem[Kim et~al.(2008)Kim, Yang, Lee, and Cho]{kim2008usefulness}
Hyun~Cheol Kim, Dal~Mo Yang, Sang~Ho Lee, and Yong~Duck Cho.
\newblock Usefulness of renal volume measurements obtained by a 3-dimensional
  sonographic transducer with matrix electronic arrays.
\newblock \emph{Journal of Ultrasound in Medicine}, 27\penalty0 (12):\penalty0
  1673--1681, 2008.

\bibitem[Korkmaz et~al.(2018)Korkmaz, Aras, G{\"u}neyli, and
  Y{\i}lmaz]{korkmaz2018clinical}
Mehmet Korkmaz, Bekir Aras, Serkan G{\"u}neyli, and M{\"u}mtaz Y{\i}lmaz.
\newblock Clinical significance of renal cortical thickness in patients with
  chronic kidney disease.
\newblock \emph{Ultrasonography}, 37\penalty0 (1):\penalty0 50, 2018.

\bibitem[Kovesdy(2022)]{kovesdy2022epidemiology}
Csaba~P Kovesdy.
\newblock Epidemiology of chronic kidney disease: an update 2022.
\newblock \emph{Kidney International Supplements}, 12\penalty0 (1):\penalty0
  7--11, 2022.

\bibitem[Maier-Hein et~al.(2020)Maier-Hein, Reinke, Kozubek, Martel, Arbel,
  Eisenmann, Hanbury, Jannin, M{\"u}ller, Onogur, et~al.]{maier2020bias}
Lena Maier-Hein, Annika Reinke, Michal Kozubek, Anne~L Martel, Tal Arbel,
  Matthias Eisenmann, Allan Hanbury, Pierre Jannin, Henning M{\"u}ller, Sinan
  Onogur, et~al.
\newblock Bias: Transparent reporting of biomedical image analysis challenges.
\newblock \emph{Medical image analysis}, 66:\penalty0 101796, 2020.

\bibitem[Marsousi et~al.(2016)Marsousi, Plataniotis, and
  Stergiopoulos]{marsousi2016automated}
Mahdi Marsousi, Konstantinos~N Plataniotis, and Stergios Stergiopoulos.
\newblock An automated approach for kidney segmentation in three-dimensional
  ultrasound images.
\newblock \emph{IEEE Journal of Biomedical and Health Informatics}, 21\penalty0
  (4):\penalty0 1079--1094, 2016.

\bibitem[Mendoza et~al.(2013)Mendoza, Kang, Safdar, Myers, Peters, and
  Linguraru]{mendoza2013kidney}
Carlos~S Mendoza, Xin Kang, Nabile Safdar, Emmarie Myers, Craig~A Peters, and
  Marius~George Linguraru.
\newblock Kidney segmentation in ultrasound via genetic initialization and
  active shape models with rotation correction.
\newblock In \emph{2013 IEEE 10th International Symposium on Biomedical
  Imaging}, pages 69--72. IEEE, 2013.

\bibitem[Netter(2018)]{netter2018atlas}
Frank~H Netter.
\newblock \emph{Atlas of Human Anatomy: Latin Terminology E-Book: English and
  Latin Edition}.
\newblock Elsevier Health Sciences, 2018.

\bibitem[Ouyang et~al.(2020)Ouyang, He, Ghorbani, Yuan, Ebinger, Langlotz,
  Heidenreich, Harrington, Liang, Ashley, et~al.]{ouyang2020video}
David Ouyang, Bryan He, Amirata Ghorbani, Neal Yuan, Joseph Ebinger, Curtis~P
  Langlotz, Paul~A Heidenreich, Robert~A Harrington, David~H Liang, Euan~A
  Ashley, et~al.
\newblock Video-based ai for beat-to-beat assessment of cardiac function.
\newblock \emph{Nature}, 580\penalty0 (7802):\penalty0 252--256, 2020.

\bibitem[Reinke et~al.(2021)Reinke, Eisenmann, Tizabi, Sudre, R{\"a}dsch,
  Antonelli, Arbel, Bakas, Cardoso, Cheplygina, et~al.]{reinke2021common}
Annika Reinke, Matthias Eisenmann, Minu~D Tizabi, Carole~H Sudre, Tim
  R{\"a}dsch, Michela Antonelli, Tal Arbel, Spyridon Bakas, M~Jorge Cardoso,
  Veronika Cheplygina, et~al.
\newblock Common limitations of image processing metrics: A picture story.
\newblock \emph{arXiv preprint arXiv:2104.05642}, 2021.

\bibitem[Remer et~al.(2014)Remer, Papanicolaou, Casalino, Bishoff, Blaufox,
  Coursey, Dighe, Eberhardt, Goldfarb, Harvin, et~al.]{remer2014acr}
Erick~M Remer, Nicholas Papanicolaou, David~D Casalino, Jay~T Bishoff, M~Donald
  Blaufox, Courtney~A Coursey, Manjiri Dighe, Steven~C Eberhardt, Stanley
  Goldfarb, Howard~J Harvin, et~al.
\newblock Acr appropriateness criteria{\textregistered} on renal failure.
\newblock \emph{The American Journal of Medicine}, 127\penalty0 (11):\penalty0
  1041--1048, 2014.

\bibitem[Rothberg et~al.(2021)Rothberg, Ralston, Rothberg, Martin, Zahorian,
  Alie, Sanchez, Chen, Chen, Thiele, et~al.]{rothberg2021ultrasound}
Jonathan~M Rothberg, Tyler~S Ralston, Alex~G Rothberg, John Martin, Jaime~S
  Zahorian, Susan~A Alie, Nevada~J Sanchez, Kailiang Chen, Chao Chen, Karl
  Thiele, et~al.
\newblock Ultrasound-on-chip platform for medical imaging, analysis, and
  collective intelligence.
\newblock \emph{Proceedings of the National Academy of Sciences}, 118\penalty0
  (27):\penalty0 e2019339118, 2021.

\bibitem[Sanusi et~al.(2009)Sanusi, Arogundade, Famurewa, Akintomide, Soyinka,
  Ojo, and Akinsola]{sanusi2009relationship}
Abubakr~A Sanusi, Fatiu~A Arogundade, Olusola~C Famurewa, Anthony~O Akintomide,
  Folashade~O Soyinka, Olalekan~E Ojo, and Adewale Akinsola.
\newblock Relationship of ultrasonographically determined kidney volume with
  measured gfr, calculated creatinine clearance and other parameters in chronic
  kidney disease (ckd).
\newblock \emph{Nephrology Dialysis Transplantation}, 24\penalty0 (5):\penalty0
  1690--1694, 2009.

\bibitem[Singla et~al.(2022)Singla, Kadatz, Rohling, and
  Nguan]{singla2022kidney}
Rohit Singla, Matthew Kadatz, Robert Rohling, and Christopher Nguan.
\newblock Kidney ultrasound for the nephrologist: A review.
\newblock \emph{Kidney Medicine}, page 100464, 2022.

\bibitem[Szabo(2004)]{szabo2004diagnostic}
Thomas~L Szabo.
\newblock \emph{Diagnostic ultrasound imaging: inside out}.
\newblock Academic press, 2004.

\bibitem[Taffel et~al.(2017)Taffel, Nikolaidis, Beland, Blaufox, Dogra,
  Goldfarb, Gore, Harvin, Heilbrun, Heller, et~al.]{taffel2017acr}
Myles~T Taffel, Paul Nikolaidis, Michael~D Beland, M~Donald Blaufox, Vikram~S
  Dogra, Stanley Goldfarb, John~L Gore, Howard~J Harvin, Marta~E Heilbrun,
  Matthew~T Heller, et~al.
\newblock Acr appropriateness criteria{\textregistered} renal transplant
  dysfunction.
\newblock \emph{Journal of the American College of Radiology}, 14\penalty0
  (5):\penalty0 S272--S281, 2017.

\bibitem[Torres et~al.(2018)Torres, Queiros, Morais, Oliveira, Fonseca, and
  Vilaca]{torres2018kidney}
Helena~R Torres, Sandro Queiros, Pedro Morais, Bruno Oliveira, Jaime~C Fonseca,
  and Joao~L Vilaca.
\newblock Kidney segmentation in ultrasound, magnetic resonance and computed
  tomography images: A systematic review.
\newblock \emph{Computer methods and programs in biomedicine}, 157:\penalty0
  49--67, 2018.

\bibitem[Widjaja et~al.(2004)Widjaja, Oxtoby, Hale, Jones, Harden, and
  McCall]{widjaja2004ultrasound}
E~Widjaja, JW~Oxtoby, TL~Hale, PW~Jones, PN~Harden, and IW~McCall.
\newblock Ultrasound measured renal length versus low dose ct volume in
  predicting single kidney glomerular filtration rate.
\newblock \emph{The British journal of radiology}, 77\penalty0 (921):\penalty0
  759--764, 2004.

\bibitem[Xie et~al.(2005)Xie, Jiang, and Tsui]{xie2005segmentation}
Jun Xie, Yifeng Jiang, and Hung-tat Tsui.
\newblock Segmentation of kidney from ultrasound images based on texture and
  shape priors.
\newblock \emph{IEEE transactions on medical imaging}, 24\penalty0
  (1):\penalty0 45--57, 2005.

\bibitem[Yin et~al.(2020)Yin, Peng, Li, Zhang, You, Fischer, Furth, Tasian, and
  Fan]{yin2020automatic}
Shi Yin, Qinmu Peng, Hongming Li, Zhengqiang Zhang, Xinge You, Katherine
  Fischer, Susan~L Furth, Gregory~E Tasian, and Yong Fan.
\newblock Automatic kidney segmentation in ultrasound images using subsequent
  boundary distance regression and pixelwise classification networks.
\newblock \emph{Medical image analysis}, 60:\penalty0 101602, 2020.

\bibitem[Zanoli et~al.(2014)Zanoli, Romano, Romano, Rastelli, Rapisarda,
  Granata, Fatuzzo, Malaguarnera, and Castellino]{zanoli2014renal}
Luca Zanoli, Giulia Romano, Marcello Romano, Stefania Rastelli, Francesco
  Rapisarda, Antonio Granata, Pasquale Fatuzzo, Mariano Malaguarnera, and
  Pietro Castellino.
\newblock Renal function and ultrasound imaging in elderly subjects.
\newblock \emph{The Scientific World Journal}, 2014, 2014.

\bibitem[Zhou et~al.(2020)Zhou, Jia, Ni, Noble, Huang, and
  Tan]{zhou2020thyroid}
J~Zhou, X~Jia, D~Ni, A~Noble, R~Huang, and T~Tan.
\newblock Thyroid nodule segmentation and classification in ultrasound images,
  2020.

\end{thebibliography}


\appendix
\section{Datasheet for the Data Set}
	In this appendix, we provide a completed data sheet for the proposed data set in efforts to increase transparency and accountability. 

\textbf{For what purpose was the data set created? Was there a specific task in mind? Was there a specific gap that needed to be filled? Please provide a description.} \\
For the purposes of multi-class kidney segmentation in ultrasound imaging. It focuses on binary segmentation of the capsule from the foreground as well as multi-class segmentation of the cortex, medulla, and central echogenic complex. The literature lacks a standard data set for fair comparisons across algorithms. \\

\textbf{Who created this data set (e.g., which team, research group) and on behalf of which entity (e.g., company, institution, organization)?}\\
The initial version was created by Rohit Singla, Cailin Ringstrom, Grace Hu, Victoria Lessoway, Janice Reid, Christopher Nguan and Robert Rohling. They are researchers at the University of British Columbia in Vancouver, British Columbia, Canada.\\

\textbf{What support was needed to make this data set? (e.g., who funded the creation of the data set? If there is an associated grant, provide the name of the grantor and the grant name and number, or if it was supported by a company or government agency, give those details.)} \\
No specific funding or grant was provided for the creation of this data set. The creators are supported by funding from the Natural Sciences and Engineering Council of Canada, the Kidney Foundation of Canada, and the American Society of Transplant Surgeons.\\

\textbf{Any other comments?}\\
n/a.\\

\textbf{What do the instances that comprise the data set represent (e.g., documents, photos, people, countries)? Are there multiple types of instances (e.g., movies, users, and ratings; people and interactions between them; nodes and edges)? Please provide a description.} \\
Each instance is an ultrasound image with a label for quality, view type, kidney type, and polygon annotations for each class. There are between 0-4 annotations for any given image. \\

\textbf{How many instances are there in total (of each type, if appropriate)?} \\
There are a 514 unique B-mode images with 20 additional copies (two sets of 10) repeated from these 514. 534 images in total. \\

\textbf{Does the data set contain all possible instances or is it a sample (not necessarily random) of instances from a larger set? If the data set is a sample, then what is the larger set? Is the sample representative of the larger set (e.g., geographic coverage)? If so, please describe how this representativeness was validated/verified. If it is not representative of the larger set, please describe why not (e.g., to cover a more diverse range of instances, because instances were withheld or unavailable).}\\
The data set does not contain all possible instances. It is a random sampling of retrospective data from a local institution, rather than a comprehensive global sampling. The sampling is representative of the Canadian population, specifically in the province of British Columbia.\\

\textbf{What data does each instance consist of?    “Raw” data (e.g., unprocessed text or images) or features? In either case, please provide a description.}\\
Each instance is an ultrasound image in PNG format. The resolution varies depending on the type of machine originally used to acquire the image. Images are randomly sampled from the original video (cine). Images have personally identifying information removed.\\

\textbf{Is there a label or target associated with each instance?  If so, please provide a description.}\\
Each image is accompanied by a label indicating quality, view type, kidney type, as well as a set of coordinates reflecting the class and polygon annotation of that class.\\

\textbf{Is any information missing from individual instances? If so, please provide a description, explaining why this information is missing (e.g., because it was unavailable). This does not include intentionally removed information, but might include, e.g., redacted text.}\\
For confidentiality and privacy reasons, personally identifiable information has been removed from the images. The images were originally stored in the hospital system in DICOM format. Conversion to PNG as well as redaction of information in the images themselves has been performed. \\

\textbf{Are relationships between individual instances made explicit (e.g., users’ movie ratings, social network links)? If so, please describe how these relationships are made explicit.}\\
There are no known relationships between instances.\\

\textbf{Are there recommended data splits (e.g., training, development/validation, testing)? If so, please provide a description of these splits, explaining the rationale behind them.}\\
The data splits used for training, validation and testing are reported. \\

\textbf{Are there any errors, sources of noise, or redundancies in the data set? If so, please provide a description.}\\
A list of errors is maintained online at \url{https://rsingla92.github.io/kidneyUS/}

There is no existing robustly validated definition of the central echogenic complex. Our definition broadly incorporates different minor parts of the kidney’s anatomy. 

While the expert sonographers are experts in ultrasound, there many still exist discrepancies between their annotations and how another expert may interpret the image. This aleatoric uncertainty is captured to a small degree in the inter-rater variance above.

Potential sources of error may include inaccurate segmentation masks or incorrect class labels. This has been mitigated through the quality assurance measures performed.\\

\textbf{Is the data set self-contained, or does it link to or otherwise rely on external resources (e.g., websites, tweets, other data sets)? If it links to or relies on external resources, a) are there guarantees that they will exist, and remain constant, over time; b) are there official archival versions of the complete data set (i.e., including the external resources as they existed at the time the data set was created); c) are there any restrictions (e.g., licenses, fees) associated with any of the external resources that might apply to a future user? Please provide descriptions of all external resources and any restrictions associated with them, as well as links or other access points, as appropriate.}\\
Self-contained.\\

\textbf{Does the data set contain data that might be considered confidential (e.g., data that is protected by legal privilege or by doctor-patient confidentiality, data that includes the content of individuals non-public communications)? If so, please provide a description.}\\
The nature of medical imaging data including ultrasound is inherently confidential. As a result, we have taken steps to anonymize the images and consulted our institution’s ethics board for approval as well as the data release management office.\\

\textbf{Does the data set contain data that, if viewed directly, might be offensive, insulting, threatening, or might otherwise cause anxiety? If so, please describe why.}\\
No. \\

\textbf{Does the data set relate to people? If not, you may skip the remaining questions in this section}\\
Yes. The data set contains images of people’s kidneys.\\

\textbf{Does the data set identify any subpopulations (e.g., by age, gender)? If so, please describe how these subpopulations are identified and provide a description of their respective distributions within the data set}\\
No. The subpopulation of native kidney versus transplant kidney may be the only potential subpopulation that arises from this data. \\

\textbf{Is it possible to identify individuals (i.e., one or more natural persons), either directly or indirectly (i.e., in combination with other data) from the data set? If so, please describe how.}\\
No. \\

\textbf{Does the data set contain data that might be considered sensitive in any way (e.g., data that reveals racial or ethnic origins, sexual orientations, religious beliefs, political opinions or union memberships, or locations; financial or health data; biometric or genetic data; forms of government identification, such as social security numbers; criminal history)? If so, please provide a description}\\
The data set contains medical imaging data that may or may not include pathology. The images may also be used to ascertain biometrics such as kidney length or width. \\

\textbf{How was the data associated with each instance acquired? Was the data directly observable (e.g., raw text, movie ratings), reported by subjects (e.g., survey responses), or indirectly inferred/derived from other data (e.g., part-of-speech tags, model-based guesses for age or language)? If data was reported by subjects or indirectly inferred/derived from other data, was the data validated/verified? If so, please describe how. }\\
After ethics board approval, the data was retrieved from our institution’s picture archiving and communications system. The data was then anonymized internally and processed to remove any potentially identifying information. The data was then reviewed by expert sonographers who generated the corresponding labels and annotations.\\

\textbf{What mechanisms or procedures were used to collect the data (e.g., hardware apparatus or sensor, manual human curation, software program, software API)? How were these mechanisms or procedures validated? }\\
Original image acquisition was performed by a variety of ultrasound machines and manufacturers. Image retrieval was performed using our institution’s picture archiving and communications system. Image anonymization was performed internally using automatic methods. Image annotation and review was performed using the VGG Image Annotator tool.\\

\textbf{If the data set is a sample from a larger set, what was the sampling strategy (e.g., deterministic, probabilistic with specific sampling probabilities)? }
Random sampling of retrospectively collected ultrasound images. The larger set includes all adult individuals who have received a kidney ultrasound for suspicion of kidney disease, as well as adult kidney transplant recipients who have received an ultrasound. From the larger set, a complete random sampling was taken. One frame per individual is used. \\

\textbf{Who was involved in the data collection process (e.g., students, crowd workers, contractors) and how were they compensated (e.g., how much were crowd workers paid)? }\\
Research assistants, surgeons, and sonographers were involved. No funding was provided. \\

\textbf{Over what timeframe was the data collected? Does this timeframe match the creation timeframe of the data associated with the instances (e.g., recent crawl of old news articles)? If not, please describe the timeframe in which the data associated with the instances was created.}\\
2014 to 2019\\

\textbf{Were any ethical review processes conducted (e.g., by an institutional review board)? If so, please provide a description of these review processes, including the outcomes, as well as a link or other access point to any supporting documentation.}\\
Yes. H21-02375\\

\textbf{Does the data set relate to people? If not, you may skip the remaining questions in this section.}\\ 
Yes.\\

\textbf{Did you collect the data from the individuals in question directly, or obtain it via third parties or other sources (e.g., websites)?}\\
The data was retrieved from our institution’s storage.\\
 
\textbf{Were the individuals in question notified about the data collection? If so, please describe (or show with screenshots or other information) how notice was provided, and provide a link or other access point to, or otherwise reproduce, the exact language of the notification itself.}\\
No.\\

\textbf{Did the individuals in question consent to the collection and use of their data? If so, please describe (or show with screenshots or other information) how consent was requested and provided, and provide a link or other access point to, or otherwise reproduce, the exact language to which the individuals consented}\\
No. In our research ethics application, given the scale of the initial retrieval and the subsequent anonymization, individuals were not directly notified.\\

\textbf{If consent was obtained, were the consenting individuals provided with a mechanism to revoke their consent in the future or for certain uses? If so, please provide a description, as well as a link or other access point to the mechanism (if appropriate).}\\
n/a \\

\textbf{Has an analysis of the potential impact of the data set and its use on data subjects (e.g., a data protection impact analysis) been conducted? If so, please provide a description of this analysis, including the outcomes, as well as a link or other access point to any supporting documentation. }\\
No formal analysis has been conducted. Our local data release management office which handles privacy and confidentiality concerns was consulted during our ethics approval process. \\

\textbf{Was any pre-processing/cleaning/labelling of the data done (e.g., discretization or bucketing, tokenization, part-of-speech tagging, SIFT feature extraction, removal of instances, processing of missing values)? If so, please provide a description. If not, you may skip the remainder of the questions in this section.}\\
Cleaning of the data was performed. DICOM to PNG conversion, meta-data/header cleaning, as well as ultrasound machine-specific removal of personally identifiable information was performed. No additional resizing or processing was performed. \\

\textbf{Was the “raw” data saved in addition to the pre-processed/cleaned/labelled data (e.g., to support unanticipated future uses)? If so, please provide a link or other access point to the “raw” data.}\\
No\\

\textbf{Is the software used to pre-process/clean/label the instances available? If so, please provide a link or other access point.}\\
All software used to process the data is available and open sourced.\\

\textbf{Any other comments?}\\
n/a\\

\textbf{Has the data set been used for any tasks already?}\\
A trained segmentation model for the kidney capsule task has been additionally used to automatically measure biometrics of kidney length and width in another set of images.\\

\textbf{Is there a repository that links to any or all papers or systems that use the data set? If so, please provide a link or other access point}
\url{https://rsingla92.github.io/kidneyUS/}

\textbf{What (other) tasks could the data set be used for?}\\
Automatic biometry, image quality assessment, artefact detection\\

\textbf{Is there anything about the composition of the data set or the way it was collected and pre-processed/cleaned/labelled that might impact future uses? For example, is there anything that a future user might need to know to avoid uses that could result in unfair treatment of individuals or groups (e.g., stereotyping, quality of service issues) or other undesirable harms (e.g., financial harms, legal risks) If so, please provide a description. Is there anything a future user could do to mitigate these undesirable harms?}\\
To the best of our knowledge, there is nothing regarding the composition that may impact future uses. \\

\textbf{Are there tasks for which the data set should not be used? If so, please provide a description.}\\
To the best of our knowledge, there are no such tasks. \\

\textbf{Will the data set be distributed to third parties outside of the entity (e.g., company, institution, organization) on behalf of which the data set was created? If so, please provide a description.}\\
Third parties outside of the creators will be permitted to access the data. This requires registration via the website, including a research use agreement. \\

\textbf{How will the data set be distributed (e.g., tarball on website, API, GitHub)? Does the data set have a digital object identifier (DOI)?}\\
The data is distributed through a Microsoft OneDrive link, which includes a zip of the images themselves as well as spreadsheets for the labels. \\

\textbf{When will the data set be distributed?}\\
The data set was first released in May 2022 as a pre-print.\\

\textbf{Will the data set be distributed under a copyright or other intellectual property (IP) license, and/or under applicable terms of use (ToU)? If so, please describe this license and/or ToU, and provide a link or other access point to, or otherwise reproduce, any relevant licensing terms or ToU, as well as any fees associated with these restrictions.}\\
Yes, the data set is distributed under a CC-BY-NC-SA license, restricting commercial usage. \\

\textbf{Have any third parties imposed IP-based or other restrictions on the data associated with the instances? If so, please describe these restrictions, and provide a link or other access point to, or otherwise reproduce, any relevant licensing terms, as well as any fees associated with these restrictions. }\\
There are no fees or restrictions.\\

\textbf{Do any export controls or other regulatory restrictions apply to the data set or to individual instances? If so, please describe these restrictions, and provide a link or other access point to, or otherwise reproduce, any supporting documentation. }\\
Unknown\\

\textbf{Who will be supporting/hosting/maintaining the data set? }
The data set is hosted at the University of British Columbia using the institution’s instance of Microsoft OneDrive. \\

\textbf{How can the owner/curator/manager of the data set be contacted (e.g., email address)?}\\
All questions and comments can be sent to Rohit Singla: rsingla@ece.ubc.ca\\

\textbf{ Is there an erratum? If so, please provide a link or other access point. }
All changes to the data set will be maintained under an erratum located at \url{https://github.com/rsingla92/kidneyUS/blob/main/README.md\#errata} \\

\textbf{Will the data set be updated (e.g., to correct labelling errors, add new instances, delete instances)? If so, please describe how often, by whom, and how updates will be communicated to users (e.g., mailing list, GitHub)? }\\
All changes to the data set will be maintained on the website as well as through the registration list.\\

\textbf{If the data set relates to people, are there applicable limits on the retention of the data associated with the instances (e.g., were individuals in question told that their data would be retained for a fixed period of time and then deleted)? If so, please describe these limits and explain how they will be enforced.}\\
 No. \\

\textbf{Will older versions of the data set continue to be supported/hosted/maintained? If so, please describe how. If not, please describe how its obsolescence will be communicated to users.}\\ 
They will continue to be supported with all information online unless otherwise communicated.\\

\textbf{If others want to extend/augment/build on/contribute to the data set, is there a mechanism for them to do so? If so, please provide a description. Will these contributions be validated/verified? If so, please describe how. If not, why not? Is there a process for communicating/distributing these contributions to other users? If so, please provide a description}\\
There currently is no such mechanism in place. \\

\textbf{Any other comments? }\\
n/a
\end{document}